\documentclass{article}
\usepackage{spconf,amsmath,graphicx}
\usepackage{epsfig,rotating,setspace,latexsym,amsmath,epsf,amssymb,bm}
\usepackage{color}

\usepackage{amsmath}
\usepackage{rotating,latexsym,amsmath,amssymb,bm}
\usepackage{cite}
\usepackage{amsthm}
\usepackage[caption=false,font=footnotesize]{subfig}
\usepackage{graphicx}

\def\nb0{{\mathbf{0}}}
\def\nb1{{\mathbf{1}}}

\def\ncalB{{\mathcal{B}}}
\def\ncalC{{\mathcal{C}}}

\def\ncalH{{\mathcal{H}}}

\def\ncalW{{\mathcal{W}}}

\def\nbbN{{\mathbb{N}}}

\def\nbbP{{\mathbb{P}}}

\def\argmin{\operatorname{arg~min}}

\def\E{\mathbb{E}}

\usepackage{bbm}

\title{Online Learning-based Waveform Selection for Improved Vehicle Recognition in Automotive Radar}
\name{Charles E. Thornton, William W. Howard, and R. Michael Buehrer\thanks{Correspondence: $thorntonc@vt.edu$.}}
\address{Wireless $@$ Virginia Tech, Bradley Department of ECE, Blacksburg, VA}
\begin{document}
%
\maketitle
\begin{abstract}
This paper describes important considerations and challenges associated with online reinforcement-learning based waveform selection for target identification in frequency modulated continuous wave (FMCW) automotive radar systems. We present a novel learning approach based on satisficing Thompson sampling, which quickly identifies a waveform expected to yield satisfactory classification performance. We demonstrate through measurement-level simulations that effective waveform selection strategies can be quickly learned, even in cases where the radar must select from a large catalog of candidate waveforms. The radar learns to adaptively select a bandwidth for appropriate resolution and a slow-time unimodular code for interference mitigation in the scene of interest by optimizing an expected classification metric.
\end{abstract}
\section{Introduction}
\label{sec:intro}
Highly autonomous vehicles will rely heavily on active sensing technology for holistic environmental perception \cite{Dickman2016,patole2017automotive}. To enable greater degrees of autonomy, next-generation active sensors, which include radar and lidar, should exhibit high-resolution imaging capabilities and maintain sufficient reliability in dynamic scenarios \cite{Hakobyan2019,engels2021automotive}. Further, to support commercial operation, the active sensors must be highly integrated at a reasonable cost \cite{waldschmidt2021automotive}. 

Considering these fundamental limitations, radar is indispensable for emerging vehicular applications due to its inherent resilience to inclement weather, extended detection range, and relatively low cost of integration. Thus, it is crucial to design radar systems which perform reliably over a broad range of physical scenarios. Cognitive radar is a paradigm which aims to achieve this end by applying artificial intelligence techniques to continually adapt transmission and reception parameters \cite{Bruggenwirth2019}. Herein, we examine the problem of cognitive waveform selection for a frequency modulated continuous wave (FMCW) automotive radar, which is commonly employed in practice due to low measurement time and simplicity of processing \cite{engels2021automotive}. Specifically, the bandwidth and slow-time code of FMCW waveforms are adapted with the goal of improving vehicle classification performance.

In a broad sense, optimal waveform selection for vehicular radar systems is a complicated topic, as there are many types and functions of automotive radar. Basic automotive radar waveform design principles are described in \cite{rohling2001waveform,rohling2008radar,patole2017automotive}. However, in practice there are often a large number of scatterers in the delay-Doppler response, the scene may change rapidly, and radar-to-radar interference is a considerable challenge due to the proliferation of FMCW technology and large consumption of time-frequency resources \cite{aydogdu2019radar}. Thus, the resolution requirements of automotive radar systems may change dramatically over time, and thus satisfactory transmission strategies must be quickly learned with limited prior information. In particular, continuous wave radar is highly desirable for automotive applications due to short measurement time and low computational cost. Here, the bandwidth and unimodular code of the FMCW waveform is adapted on a frame-by-frame (equivalently CPI-to-CPI) basis.

In this contribution, we describe a general framework for the application of online reinforcement learning algorithms in a vehicular target classification setting and propose a scalable, time-sensitive approach which results in rapid learning. In general, challenges include efficiently representing the high-dimensional scene, developing an adequate yet computable classification-based loss function, and algorithm parameter selection \cite{thornton2020deep,thornton2021constrained,liu2020decentralized}. Herein, we focus on real-time design of FMCW waveforms to mitigate interference with other radars and achieve a balance of resolution and reliability for multi-target recognition and classification. 

In order to learn an effective waveform from a large number of candidates we apply the decision-theoretic principle of \emph{satisficing}. Specifically, a satisficing Thompson Sampling (sTS) algorithm is used to quickly identify a satisfactory waveform which is expected to improve classification performance. It is shown that the proposed sTS algorithm allows the radar to learn an effective waveform for classification in a more timely manner than a conventional multi-armed bandit formulation. In the context of automotive radar, this is expected to yield large improvements in classification performance, which is an inherently time-sensitive application.

\section{System Model}
\label{sec:format}
We consider an automotive radar which transmits a length $N$ sequence of FMCW chirps each frame (or CPI). The transmitted pulse-train is expressed by the sum
\begin{equation}
	s(t)=\sqrt{P_{t \mathrm{x}}} \textstyle \sum_{k=0}^{N-1} A_{k} c(t-k T),
	\label{eq:sig}
\end{equation}
where $A_{k} \in \{-1,1\}$ is an amplitude, $P_{t \mathrm{x}}$ is the transmit power, which is assumed here to be normalized, $k \in \mathbb{N}_{+}$ is the chirp index within the frame, $T$ is the chirp duration, $N$ is the number of chirps per frame, and 
\begin{equation}
c(t)=\exp \left(j 2 \pi \left(f_c+\frac{B t}{2 T} \right) t \right)
\label{eq:chirp}
\end{equation}
is the frequency modulation function for each chirp. In (\ref{eq:chirp}), $B$ is the sweep bandwidth. As seen in (\ref{eq:sig}), the radar employs a slow-time unimodular code specified by $\{A_{k}\}$. Specifically, at pulse $k$ of the CPI, the radar transmits chirp $A_{k} c(t-kT)$, where $A_{k} \in \{-1,1\}$. The purpose of this code is to mitigate interference due to neighboring radars using similar transmission parameters \cite{Tang2018,bose2022waveform}.

\begin{figure*}[t]
	\centering
	\subfloat[]{\includegraphics[scale=0.37]{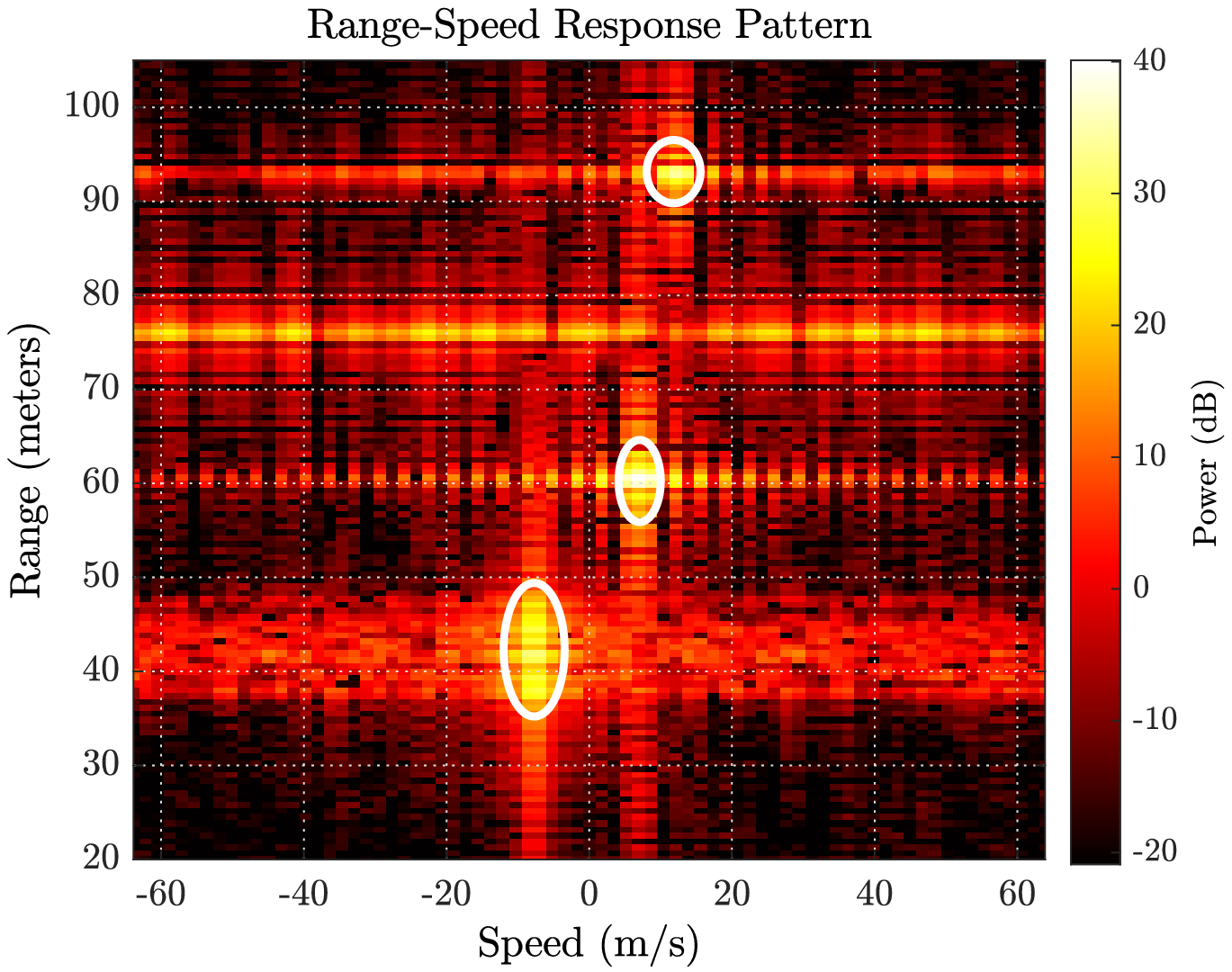}}
	\subfloat[]{\includegraphics[scale=0.37]{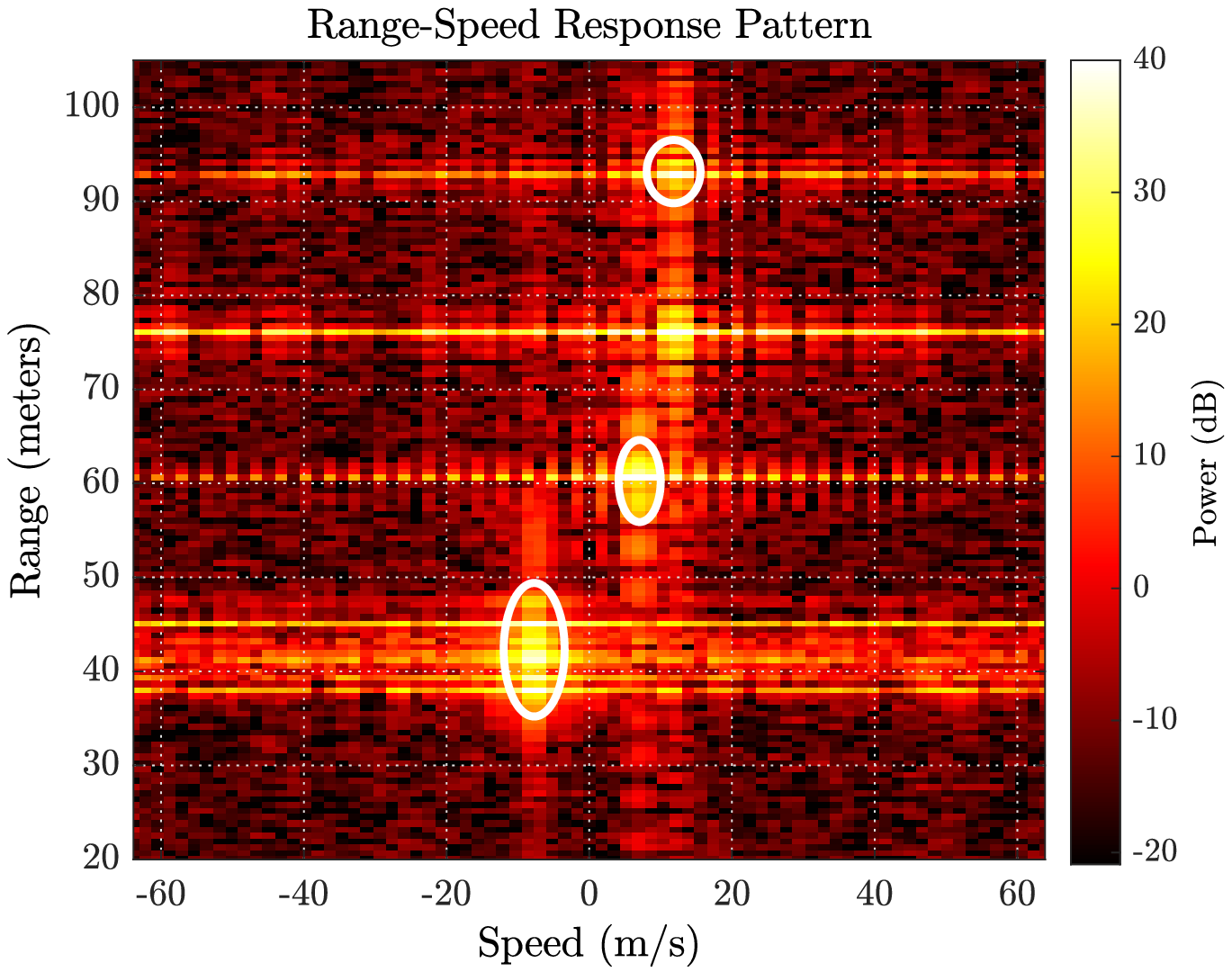}} \\
	\subfloat[]{\includegraphics[scale=0.37]{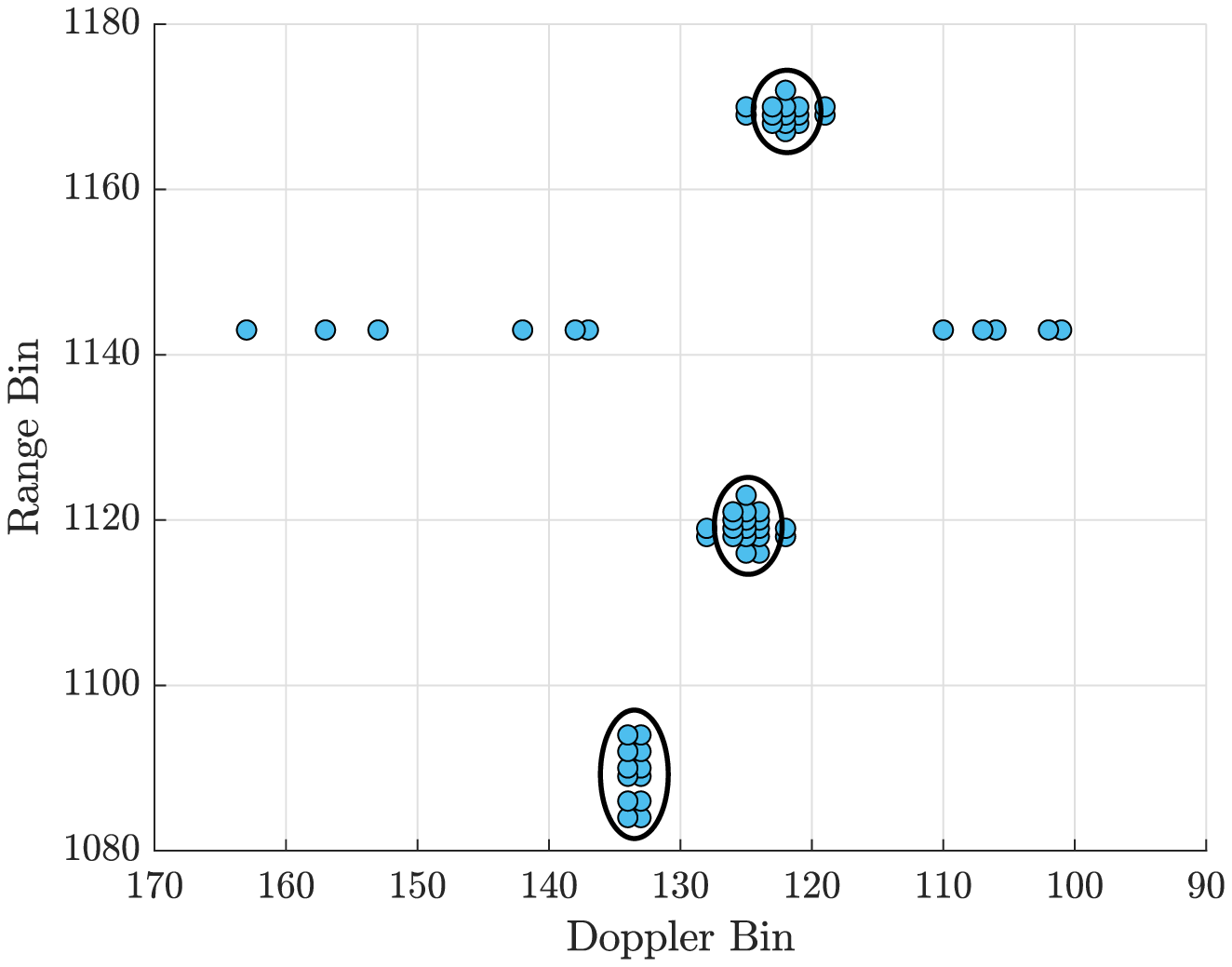}}
	\subfloat[]{\includegraphics[scale=0.37]{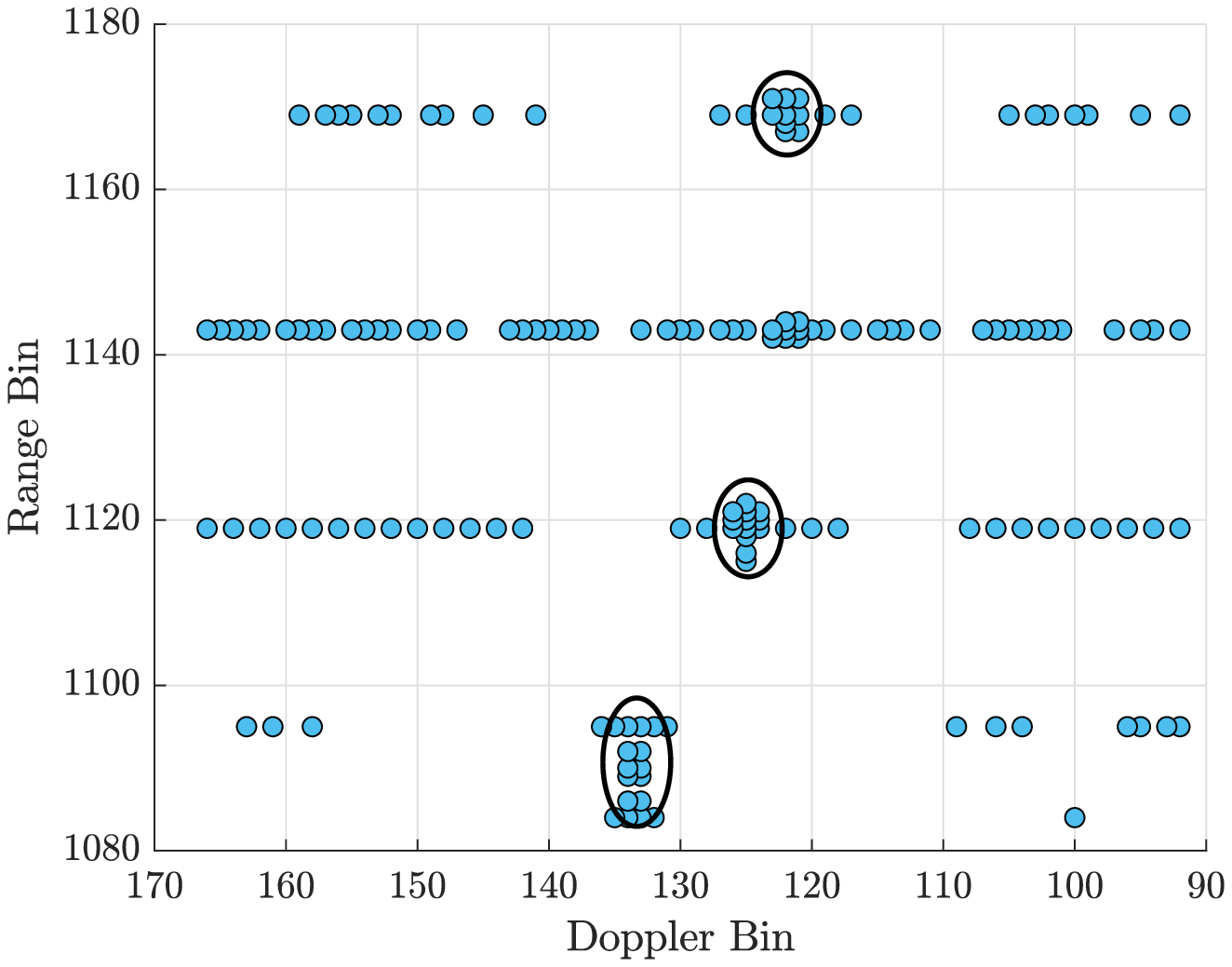}}
	\caption{Impact of slow-time coding and bandwidth adaptation on the range-Doppler images is shown in (a) and (b). Impact of waveform adaptation on CFAR-processed point-cloud data is shown in (c) and (d). True targets are located in white circles in the range-Doppler images, and in black circles in the point-cloud data.}
	\label{fig:rdData}
\end{figure*}

We assume the radar scene potentially consists of many scatterers, and may contain returns due to an interfering interfering radar. The received signal is then given by $r(t) = s_{T}(t) + s_{I}(t) + n(t)$, where $s_{T}(t)$ is the return due to target reflections, $s_{I}(t)$ is a return due to interfering signals, and $n(t)$ is signal-independent additive noise, assumed here to be stationary and Gaussian. We express the first two terms corresponding to the target and interference returns by
\begin{align}
	s_{T}(t) &= \textstyle \sum_{i=1}^{N_{T}} \alpha_{i} s(t-\tau_{i}) \exp(j 2 \pi f_{d,i}t) \\
	s_{I}(t) &= \textstyle \sum_{j=1}^{N_{I}} \alpha_{j} s(t-\tau_{j}) \exp(j 2 \pi f_{d,j}t),
\end{align}
where $\alpha_{i}$ and $\alpha_{j}$ are scalar amplitudes due to the reflection from target component $i$ and the incoming signal from interference component $j$ respectively. In general, due to the one-way propagation of the interfering signal, $\alpha_{j} > \alpha_{i}$. Thus, true targets in the range-Doppler response can easily be masked by interfering signals. These scattering coefficients can be computed using the radar range equation, but are assumed to be unknown to the radar \emph{a priori}. Similarly, $\{\tau_{i}, \tau_{j} \}$ are round trip delays, and $\{f_{d,i}, f_{d,j} \}$ are observed Doppler shifts due to target component $i$ and interference component $j$, respectively. To maintain a general model, $N_{T}$ does not necessarily equal $N_{I}$. Since automotive radars generally use large bandwidths to achieve high-resolution, 

In \cite{Tang2018,bose2022waveform}, it is shown that slow-time codes can be used to dramatically mitigate mutual interference. A simple Doppler-shifting scheme is proposed, along with a procedure for finding the optimal code. However, the optimization problem involves a joint optimization of two radars, and the problem is difficult to solve in general. Here, we focus on the selection of a code for a single radar from a pre-defined set. Notably, sub-optimal slow-time codes can be effective in mitigating ghost targets due to interference.

The received signal $r(t)$ is then decoded and de-chirped using the standard Fourier-based procedure described in \cite[Sec. II]{engels2021automotive} to produce a range-Doppler response. The range-Doppler response is then passed through a constant false-alarm rate (CFAR) detector and a point cloud is received, which can serve as an input to a clustering algorithm, such as DBSCAN \cite{ester1996density} and be used to classify objects in the radar's field of vision.

\begin{figure*}[t]
	\centering
	\subfloat[]{\includegraphics[scale=0.45]{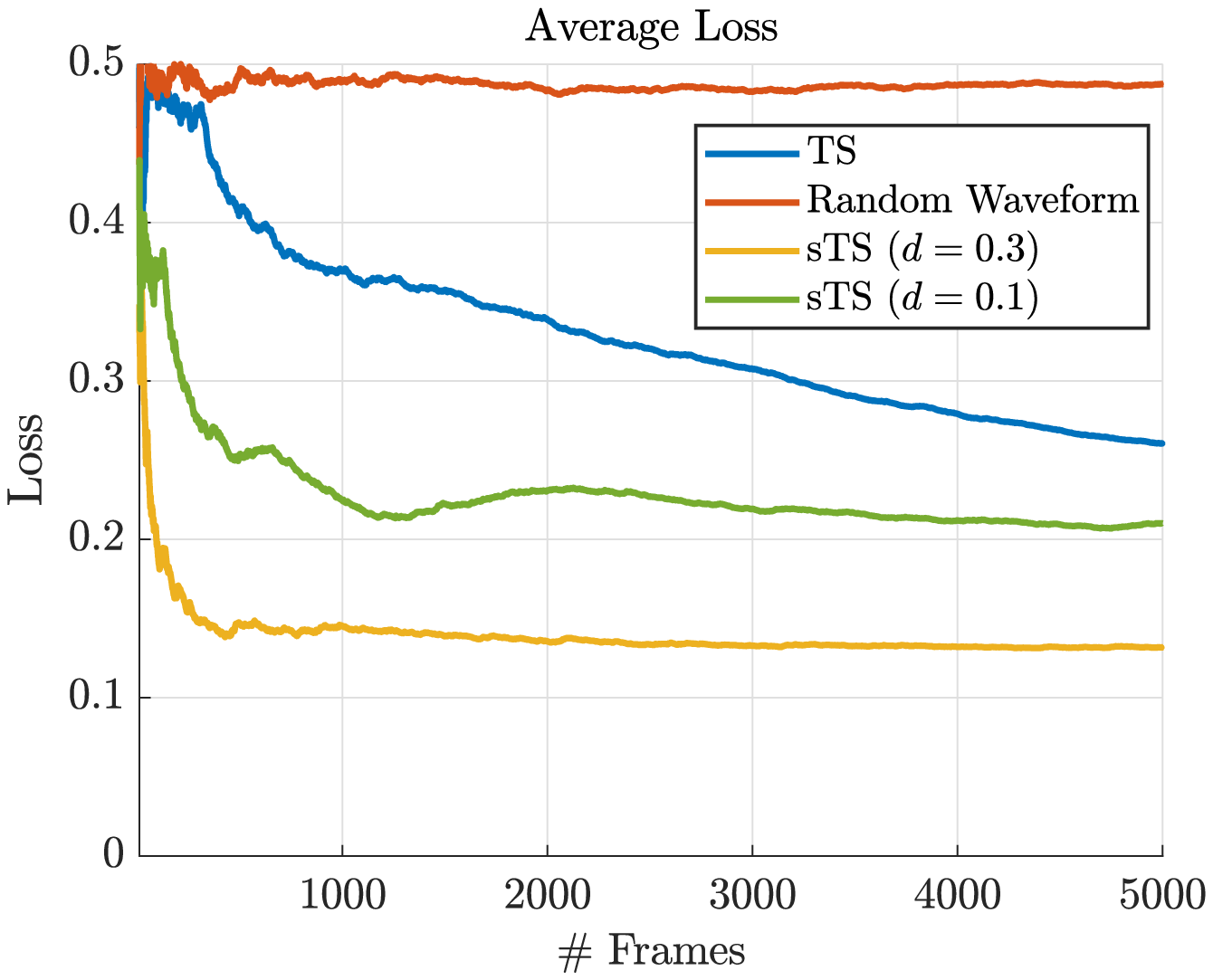}}
	\subfloat[]{\includegraphics[scale=0.45]{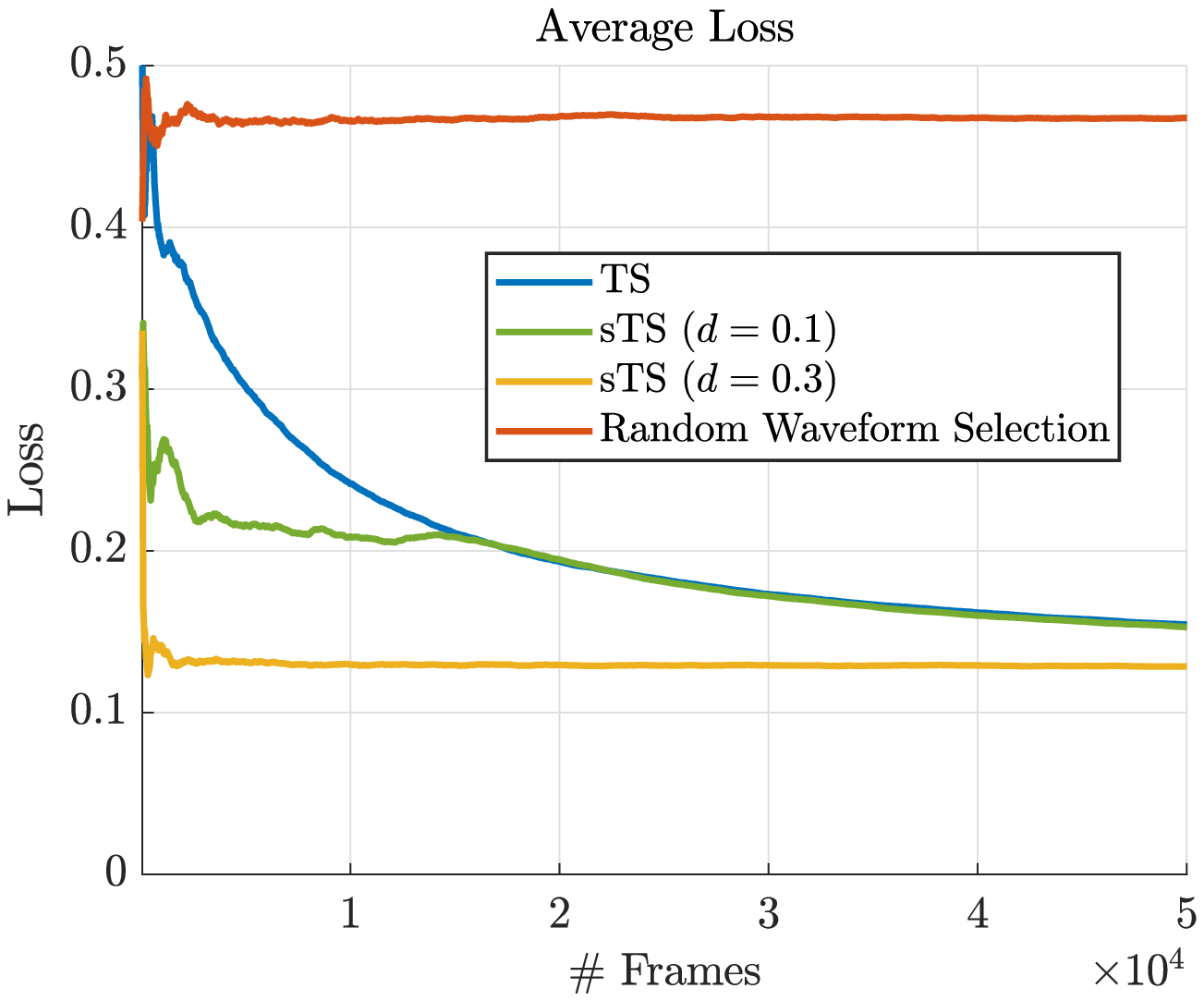}}
	\caption{Average loss of waveform selection strategies, corresponding to expected mis-classification performance. The proposed satisficing TS algorithm is employed with two distortion constraints, $d=0.1$ and $d=0.3$, and is compared to standard TS and random waveform selection.}
	\label{fig:learnCost}
\end{figure*}

\section{Learning Formulation}
\label{sec:refs}
Let the radar's waveform library during frame $t \in \mathbb{N}_{+}$ be given by set $\mathcal{W}_{t} = \{\ncalB_{t}, \ncalC_{t} \}$, where $\ncalB$ is a set of allowable bandwidths, determined by spectrum regulations and hardware constraints, and $\ncalC$ is a set containing unimodular codes of length $N$. To evaluate performance, we define a clustering and classification function $g: \mathbb{C}^{N_{r} \times N_{d}} \mapsto \mathbb{R}^{N_{c}}$, which takes as an input a range-Doppler response or CFAR point cloud with $N_{r}$ range cells and $N_{d}$ Doppler cells, and outputs a real-valued confidence metric corresponding to expected classification performance for each identified cluster. To simplify the learning process we assume $g \in [0,1]$ without loss of generality. Here, we assume a simple structure for $g(\cdot)$, which uses the DBSCAN clustering algorithm \cite{ester1996density} to split the CFAR detections into clusters and then performs classification using hand-crafted features, such as the expected spatial extent and velocity spread of certain types of targets. 

\begin{figure}[t]
	\centering
	\includegraphics[scale=0.5]{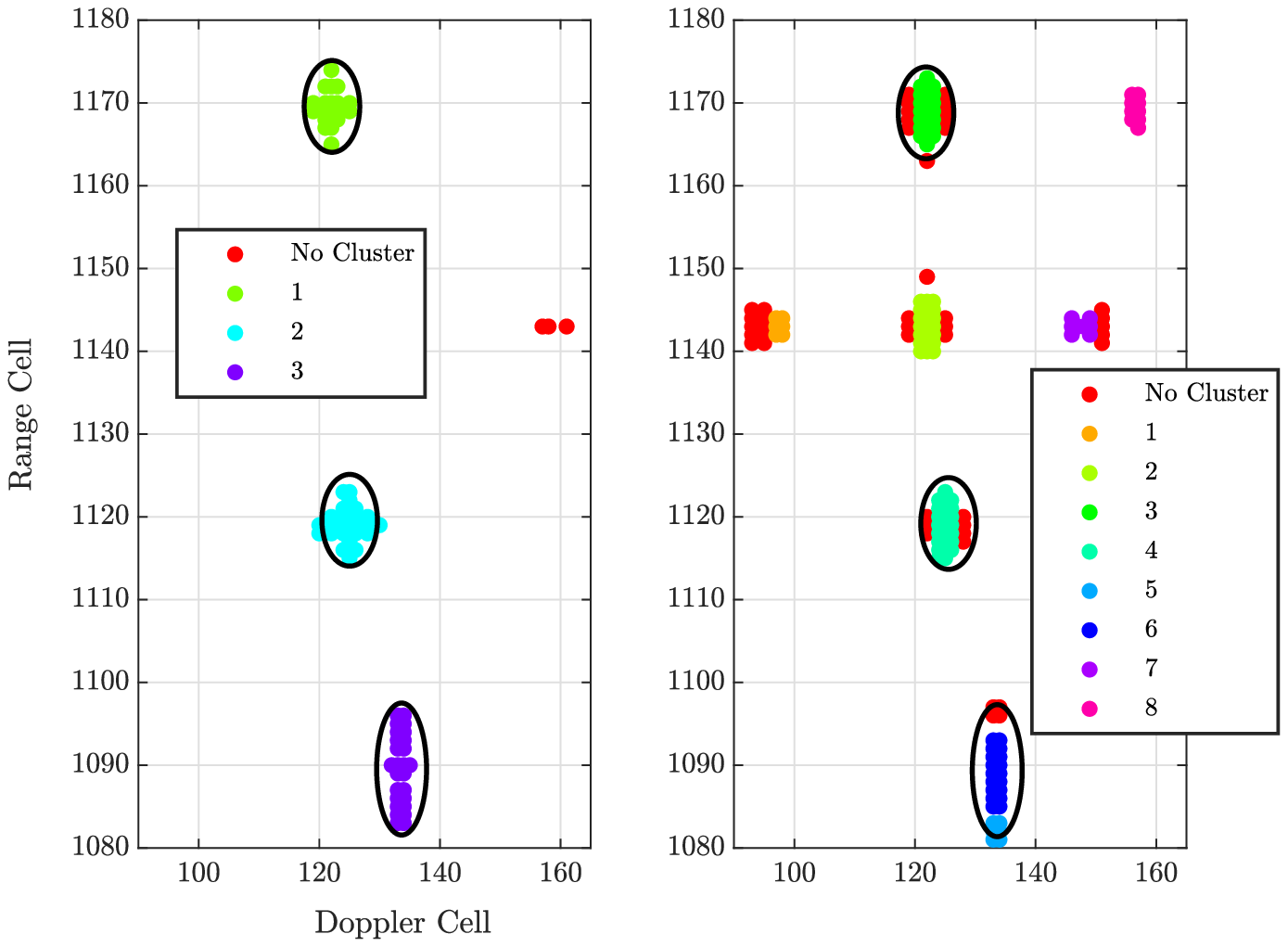}
	\caption{Impact of bandwidth and code selection on clustering performance when DBSCAN algorithm \cite{ester1996density} is used. Scene contains three true targets, located in black circles.}
\end{figure}

Each frame, the radar transmits waveform $w_{t} \in \ncalW_{t}$ and receives a loss $\ell_{t} \in [0,1]$, which is calculated using information from the range-Doppler response. Let the loss associated with transmitting waveform $w_{t}$ at frame $t$ be denoted by $\ell_{t} =  \ell(w_{t})$. We assume the radar is able to store the history of waveforms and losses up to the current frame in memory, which is given by set $\ncalH_{t-1} = \{w_{1},\ell_{1},w_{2},\ell_{2},...,w_{t-1},\ell_{t-1} \}$. The radar aims to use the history $\ncalH_{t-1}$ to select a waveform which yields a low expected cost. Here, we take the cost to be an average over the expected classification performance for each of the $N_{c}$ identified clusters, $\textstyle 1/ N_{c} \sum_{p=0}^{N_{c}} g[p]$. Let the average loss associated with each waveform $w_{i}$ be given by parameter $\theta_{w_{i}}$. Here, the radar attempts to learn a posterior distribution for each waveform $\mathbb{P}(\theta_{w_{i}})$ and transmits waveforms according to this posterior belief. This is a well-known principle in online decision problems, known as \emph{Thompson Sampling} \cite{agrawal2017near}.

Since bandwidth is a continuous value, and the number of possible slow-time codes grows exponentially with frame length $N$, the waveform catalog $\ncalW$ may be very large. MAB problems with a large number of arms are notoriously difficult to solve, since popular algorithms such as variants of the upper confidence bound principle require each arm to be tested at least once. In order to learn a near-optimal waveform from a large set of candidates over a limited time horizon, the waveform selection process is managed by a satisficing-Thompson Sampling (sTS) algorithm \cite{russo2022satisficing}. This algorithm attempts to find a \emph{satisfactory} waveform $\tilde{w} \in \mathcal{W}$, given by
\begin{equation}
	\tilde{w} = \argmin_{w_{i} \in \ncalW_{t}} \{w_{i} | \E[C(w_{i}) \geq \ell^{*} - d]\},
\end{equation}
where $\ell^{*} = \min_{w_{i}} \ell(w_{i})$ is the minimum cost achievable by the algorithm at the current frame and $d \in [0,1)$ is a distortion parameter which must be specified as an input to the algorithm. Let the average loss for waveform $w_{i}$ be specified by $\theta_{w_{i}}$. The satisficing Thompson Sampling algorithm then proceeds as follows:

\begin{enumerate}
	\item For each waveform $w \in \mathcal{W}_{t}$, sample $\theta_{w} \sim \nbbP(\theta_{w}|\ncalH_{t-1})$
	\item Let $\hat{\tau} = \min\{\tau \in \nbbN_{+}: \theta_{\tau} \geq \ell^{*} - d\}$
	\item If $\hat{\tau}$ is not the null set, transmit $w_{\hat{\tau}}$. Otherwise, transmit an untested waveform. If every waveform has been tested, transmit waveform with lowest expected loss.
	\item From range-Doppler response, observe $g_{t}$ and $\ell_{t}$. Then update the history $\ncalH_{t}$ and posterior distributions $\nbbP(\theta_{w}|\ncalH_{t})$.
\end{enumerate}

Thus, the sTS algorithm attempts to find the first waveform which is expected to yield a loss within a distance $d$ of the best possible waveform. When $d = 0$, the above algorithm is equivalent to standard Thompson Sampling. The computational efficiency of this algorithm mirrors standard Thompson Sampling implementations, which are feasible in real-time. Since the loss here is a continuous value, the sTS algorithm can be simply implemented using Gaussian TS, described in \cite[Algorithm 2]{agrawal2017near}. Further, we shall see that for waveform selection problems with a large waveform catalog, sTS results in superior short-term performance as compared to standard TS, and nearly equivalent long-term performance. 

\section{Numerical Results}
In these simulations, we consider a radar operating at $f_{c} = 77$ GHz. The radar uses an FMCW waveform, and each frame consists of $N = 128$ FMCW chirps. The scene consists of three targets, randomly distributed in the delay-Doppler space. The radar's maximum detectable range is $120$m, which corresponds to a measurement time of $4.4 \times 10^{-6}$ sec. The radar's bandwidth is selected from library $\ncalB$, which contains a discrete set of values between $30$MHz and $1.5$GHz. The unimodular code is selected from set $\ncalC$, which consists of binary-valued sequences of length $N = 128$.

In Figure \ref{fig:rdData}, we observe the impact of simultaneous bandwidth adaptation and code selection on the range-Doppler processed data. Additionally, we observe point cloud data consisting of constant false alarm rate (CFAR) detections when the probability of false alarm is set to $P_{fa} = 10^{-5}$, $5$ guard cells, and $10$ training cells are used to implement the detector. We observe that when a near-optimal choice of bandwidth and a reasonable unimodular code sequence is selected, many false detections are eliminated. In particular, a scattering return which appears due to an interfering radar at a range of $76$m and relative speed of $12$m/s is substantially mitigated.

In Figure \ref{fig:learnCost}, the average cost achieved by the proposed sTS algorithm is compared to random waveform selection and a standard Thompson Sampling approach, which is equivalent to sTS with $d = 0$. We observe that with a good choice of distortion parameter, namely $d = 0.3$, both near-term and long-term performance is superior to random waveform selection and standard TS. Further, even when a sub-optimal distortion parameter is selected, near-term behavior exceeds that of standard TS. In the long-term, namely as the number of frames approaches $5\times 10^{4}$, we observe that the expected classification performance of the sTS and standard TS approaches are roughly equivalent. Thus, we observe some degree of robustness for the proposed sTS approach in cases where the waveform library is very large.

\section{Conclusion}
This paper has introduced a computationally feasible online reinforcement learning formulation for cognitive vehicular radar based on the decision-theoretic principle of satisficing. A learning problem was defined to adapt the bandwidth and slow-time code of an FMCW waveform on a frame-by-frame basis. It was shown that satisficing Thompson Sampling can be used to select a near-optimal waveform online while requiring very few frames of exploration, thus markedly improving the probability of correct classification of moving objects. 

Future work will focus on modeling additional aspects of the scene and exploring the learnability of specific classifiers, such as deep-learning-based approaches \cite{scheiner2020off,patel2019deep}, in more detail. We note that the learning framework presented here is expected to maintain good performance in a variety of vehicle classification/recognition tasks. However, considerations related to the specific application may influence the choice of learning parameters. For example, in simple detection problems, few waveforms may be needed to populate $\mathcal{W}$, while classification problems demanding high-resolution may require $\mathcal{W}$ to be large. The nature of the waveform catalog influences the choice of tolerance parameter $d$ in the learning algorithm \cite{arumugam2021deciding}. \newpage


\bibliographystyle{IEEEbib}
\bibliography{autoBib.bib}

\end{document}